\documentclass[twocolumn,showpacs,preprintnumbers,amsmath,amssymb]{revtex4}
\usepackage[dvips]{graphicx}
\usepackage{dcolumn}
\usepackage{bm}

\newcommand{\nc}{\newcommand}
\def\nn{\nonumber\\}
\def\bea{\begin{eqnarray}}
\def\eea{\end{eqnarray}}
    
\nc{\braket}[1]{\langle\,{#1}\rangle}

\def\pa{\partial}

\def\Z{\mathbf Z}    
\def\VD{{\cal V\! D}}

\begin{document}

\title{Analytic theory of DNA condensation}

\author{Y. Ishimoto}
\affiliation{Theoretical Physics Laboratory, RIKEN, Wako 351-0198, Japan}
\author{N. Kikuchi}
\affiliation{Institut f\"ur Physik, Johannes Gutenberg-Universit\"at Mainz, Staudinger Weg 7, D-55099 Mainz, Germany}

\date{\today}

\begin{abstract}
We introduce a novel model for DNA
condensation (whip-toroid transition) using the path integral method
in the framework of the non-linear sigma model on a line segment. We show
that some of its classical configurations exhibit toroidal forms, and 
the system has phase transitions from a whip to toroidal phases
with a parameter $c=\frac{W}{2 l} \left( \frac{L}{2\pi}
\right)^2$. We also discuss stability and finite size effect on these
states.
\end{abstract}

\pacs{87.14.Gg, 87.10.+e, 64.70.Nd, 82.35.Lr}

\maketitle


Since in living cells DNA is often packaged tightly, 
for instance, inside phage capsids, 
DNA condensation has drawn much attention 
\cite{GS76,B91,B96,YYK98,CVH03,HD01} 
as well as its biochemical/medical importance in the emerging field of gene  therapy.
In fact, when we put condensing agents as multivalent cations into DNA
solution, it can cause DNA to undergo the condensation from a
worm-like chain (whip) to toroidal states 
\cite{GS76,B91,B96,YYK98}. 

A double stranded DNA chain can be modelled, for example, by a semiflexible homopolymer chain 
\cite{DE86,K04,KF72}.
To increase our understanding of ``whip-toroid transition'',
semiflexible homopolymers in a poor solvent condition (i.e effective
interactions between polymer segments are attractive) have been 
investigated as simple
models 
\cite{MPM04,CW04,NSKY96,NY98,KT99,KTD96,IPB98,GK81,SIGPB03,HDB95,UO96,SMW00,SGM02,PW00,MKPW05}.
Simulations using Monte Carlo, Langevin approaches or Gaussian variational method, 
calculated phase diagram for the semiflexible chain in a poor solvent 
\cite{MPM04,CW04,NSKY96,NY98,KT99,KTD96,IPB98}.
In theoretical works, existing phenomenological models balance the 
bending and surface free
energies to estimate toroidal properties 
\cite{GK81,SIGPB03,HDB95,UO96,SMW00,SGM02,PW00,MKPW05}. 
It becomes increasingly probable that toroid is the stable lowest
energy state --- the ground state.
We note,
however, that the theoretical aspects of the works assume a priori toroidal geometry as the stable lowest energy state with no theoretical proof 
\cite{SMW00}. 
Moreover, compared to the theory of coil-globule 
transition of flexible chains \cite{LGK78,KF84,DG79,DE86,GK94}, 
which are well described by field theoretical formalism 
\cite{LGK78,KF84}, 
there is no simple microscopic theory, which contains the salient physics to
demonstrate the whip-toroid transition of the semiflexible polymer. 
Difficulties in formulating theory results specifically from the local 
inextensibility constraint of the semiflexible chain \cite{K04,KF72}, which makes the theory non-Gaussian, and also from the non-local nature of the attractive interaction along the polymer chain, which makes the theory analytically 
intractable.

In this letter, we investigate the whip-toroid
transition of a single semiflexible homopolymer chain with attraction, 
using path integral method and the nonlinear sigma model on
the line segment.

In the continuum limit, the Green function (end-to-end distribution) of a stiff polymer chain
with attractive interactions can be given by the path
integral 
\bea
G(\vec{0},\vec{R};\vec{u}_i,\vec{u}_f;L,W)=
{\cal N}^{-1}\!\!\!\! \int_{\vec{r}(0)=\vec{0}, {\vec{u}(0)=\vec{u}_i}}^{\vec{r}(L)=\vec{R}, {\vec{u}(L)=\vec{u}_f}} 
  \!\!\!\!\!\!\!\!\!\!\!\!\!\!\!\!\!\!\!\!\!\!\!\!\! {\cal D} [\vec{r}(s)]\, e^{- {\cal H}[\vec{r}, \vec{u}, W]}
\eea
with a constraint $|\vec{u}|^2 = 1$ 
\cite{K04,KF72}. 
$s$ is the proper time along the stiff polymer chain of total contour length $L$. $\vec{r}(s)$ denotes the pointing vector at the `time' $s$ in our three dimensional space while
$\vec{u}(s)\equiv\frac{\pa \vec{r}(s)}{\pa s}$ corresponds to the unit
orientation vector at $s$. ${\cal N}$ is the normalisation
constant. Following Freed et al.
\cite{KF84,K04}, the dimensionless Hamiltonian can be
written by ${\cal H}[\vec{r}, \vec{u}, W] =
\int_0^L ds\, \left[ H(s) +
  V_{AT}(s) \right]$ where $H(s)$ and $V_{AT}(s)$ are the local free Hamiltonian and the
attractive interaction term, respectively: 
\bea
H(s) &=& \frac{l}{2} \left\vert \frac{\pa}{\pa s} \vec{u}(s)
\right\vert^2\!\!\!,\\
V_{AT}(s) &=& - W \int_0^s d s^\prime \delta \left( \vec{r}(s) - \vec{r}(s^\prime) \right)\!.\label{VDWhamiltonian}
\eea
$l$ is the persistence length and $W$ is a positive coupling constant
of the attractive interaction between polymer segments. Thermodynamic $\beta=1/(k_BT)$ is implicitly included in $l$ and $W$. 
This will be revived when we consider the
thermodynamic behaviours of the system. In what follows, we express
$\vec{r}$ by the unit tangent vector $\vec{u}$ and therefore 
the Hamiltonian ${\cal H}(\vec{u})$ in terms of 
$\vec{u}$. Hence, the Green function $G\left( \vec{0}, \vec{R};
\vec{u}_i, \vec{u}_f; L, W \right)$ becomes a path integral over $\vec{u}$ with a positive coupling constant $W$, regardless of $\vec{r}$,
\bea
G = \int_{\vec{u}_i}^{\vec{u}_f} {\cal D} [\vec{u}(s)]\, 
    \delta\left({\textstyle \int_0^L ds\, \vec{u}(s) - \vec{R} }\right) \, e^{- {\cal H}[\vec{u}, W]} ,
\eea
where we used $\vec{r}(L)= \int_0^L ds\, \vec{u}(s)$ 
and the Jacobian is absorbed by ${\cal N}$ which is neglected here. 
The delta
function selects out the end-to-end vector.
When $W=0$, our free dimensionless Hamiltonian with the constraint
$|\vec{u}(s)|^2=1$ can be interpreted as the low energy limit of a
linear sigma model on a line segment, or quantum equivalently 
a nonlinear sigma model on a 
line segment, rather than some constrained Hamiltonian system. 


We begin with $O(3)$ nonlinear sigma model on a line
segment which is nothing but a quantum mechanics of a limited time 
$s\in [0,L]$ with a constraint. 
The constraint $|\vec{u}|^2 = 1$ restricts the value of $\vec{u}$ on a 
unit sphere $S^2$. Its polar coordinate decomposition in the
free Hamiltonian ${\cal H}(\vec{u})=\frac{l}{2} \int_0^L ds\, \left| \pa \vec{u}(s) \right|^2$ gives the action: 
\bea
 S[\theta_u, \varphi_u] &=& \frac{l}{2} \int_0^L ds\; \left[ (\pa \theta_u)^2 + \sin^2 \theta_u (\pa \varphi_u)^2 \right]. 
\label{nonlinearsigmaaction}
\eea
This is called the nonlinear sigma model since the action is $O(3)$
symmetric but some of its transformations are realised nonlinearly.
Minimizing the action (\ref{nonlinearsigmaaction}) in terms of $\theta_u$
and $\varphi_u$ yields the classical equations of motion: 
\bea
 \left[ - \pa^2  + \frac{\sin 2\theta_u}{2 \theta_u} (\pa \varphi_u)^2 \right] \theta_u &=& 0,
 \nn
 \left[ \,\pa^2 + 2 (\pa \theta_u) \cot \theta_u \pa\, \right] \varphi_u &=& 0 .
\label{EOM}
\eea

Our aim is now to explore classical solutions of eq.(\ref{EOM}) and to study the whip-toroid phase transition in the presence of attractive interactions.
Consider classical solutions of eq.(\ref{EOM}) with a trial solution
$\dot{\theta_u}=0$. The first equation of (\ref{EOM}) leads to $\sin 2\theta_u (\dot{\varphi_u})^2 =0 $.
Thus the solution is either $\theta_u = 0, \frac{\pi}{2}, \pi$ or
$\dot{\varphi_u} = 0$. The solutions $\theta_u=0,\pi$ or
$\dot{\varphi_u}=0$ with $\dot{\theta_u}=0$ are equivalent to having a
constant $\vec{u}$. Accordingly, classical solutions reduce to $\theta_u =
\frac{\pi}{2}$ or $\vec{u}=const$. Substitution of $\theta_u =
\frac{\pi}{2}$ into the second equation of motion (\ref{EOM}) gives $\pa^2 \varphi_u =0$.
Therefore, we have two types of classical solutions: 
\bea
  && \vec{u}(s) = const.
  \nn
  &&or~~
  \nn
  && \theta= \frac{\pi}{2} \quad and \quad \varphi_u = a s + b ,
  \label{classical sol}
\eea
where $a, b$ are constants. 
By symmetry argument, we state that the solutions 
(\ref{classical sol}) represent the general solutions \cite{IK05}. That is either
a constant $\vec{u}(s)$ (rod solution) or a rotation at a constant speed along a
great circle (toroid solution). 

Now we consider the attractive interaction term (\ref{VDWhamiltonian}). 
It is difficult to interpret it in the context of quantum theory
due to its non-local nature along the polymer chain. 
However, we can solve them with our classical solutions (\ref{classical sol}).
Let us rewrite
eq.(\ref{VDWhamiltonian}) by $\vec{r}(s) - \vec{r}(s^\prime) =
\int_0^s dt\, \vec{u}(t) - \int_0^{s^\prime} dt\, \vec{u}(t) =
\int_{s^\prime}^s dt\, \vec{u}(t)$, that is, 
$V_{AT}(s) = -W \int_0^s d s^\prime\; \!\!\delta\left( \int_{s^\prime}^s dt\, \vec{u}(t) \right)$. Hence the problem is now reduced to the one in the $\vec{u}$ space: finding non-zero values of
$\delta\left( \int_{s^\prime}^s dt\, \vec{u}(t) \right)$ with the
classical solutions (\ref{classical sol}). 
That is to find $\vec{u}(s^\prime)$ for a given $s$, which satisfies $\int_{s^\prime}^s dt\, \vec{u} = 0$ . 
In the polar coordinates, this is expressed by
$
\int_{s^\prime}^s dt\, \sin \theta_u \cos \varphi_u = 0, 
\int_{s^\prime}^s dt\, \sin \theta_u \sin \varphi_u = 0$, 
and
$\int_{s^\prime}^s dt\, \cos \theta_u = 0. 
$
The first classical solution does not satisfy these equations and thus derives no attractive interactions. If we
substitute the second classical solution of eq.(\ref{classical sol})
into the equations, we have $\cos \theta_u(s)=0$, $\int_{s^\prime}^s
dt\, \cos (at +b) = \frac1a \left( \sin (as+b) - \sin (as^\prime+b)
\right) = 0$ and $\int_{s^\prime}^s dt\, \sin (at +b) = \frac1a \left(
\cos (as^\prime+b) - \cos (as+b)\right) = 0$. 
Hence we have solutions: $s-s^\prime = 2n\pi/a >0$, $n\in\Z$. 
Without any loss of generality, 
we assume $a>0$ and $n\in\Z_+$. Introducing $N(s) \equiv [as/2\pi]$ by Gauss' symbol
\footnote{
Gauss' symbol $[x]$ gives the greatest integer that is not exceeding $x$.
}, we obtain $\int_{s-2\pi/a}^s dt\, \vec{u}(t) = \int_{s-4\pi/a}^s
dt\, \vec{u}(t) = \cdots = \int_{s- 2\pi N(s)/a}^s dt\, \vec{u}(t) =
0$. Therefore, the attractive potential is given by 
$V_{AT}(s) = - W \cdot N(s)$. 
Note that $N(L)$ represents the winding number of the classical solution
(\ref{classical sol}) along a great circle of $S^2$. 
Finally, an integration over $s$ yields the dimensionless Hamiltonian 
with our classical solutions:
\bea
  &{\cal H}&[\vec{u},W]
  = \int_0^L ds\, H(s) + \int_0^L ds\,V_{AT}(s)
  \nn
  &=& \frac{Ll}{2} a^2 - W \left[ \frac{2\pi}{a} \sum_{k=1}^{N(L)-1} k + \frac{2\pi}{a} \left( \frac{aL}{2\pi} - N(L) \right) N(L) \right]
  \nn
   &=& \frac{Ll}{2} a^2
   - W L\cdot N(L) \left\{ 1-\frac{\pi}{aL}\left(N(L)+1\right) \right\}\label{Hamilwithattraction}
   .
\eea
The first term denotes the bending energy, 
and the second and the third terms are thought of as 
`topological' terms from the winding number.

The non-zero winding number of the classical solution in the $\vec{u}$
space means that the polymer chain winds in the $\vec{r}$ space as
well. That is, when $a>\frac{2\pi}{L}$, configurations around the
second classical solution (\ref{classical sol}) start forming a toroidal shape since
\bea
 \vec{r}(s) = \left( \begin{array}{c} 
    \frac{1}{a} \left\{ \sin (as +b)-\sin (b) \right\} \\
    -\frac{1}{a} \left\{ \cos (as +b)-\cos (b) \right\} \\
    const.
  \end{array} \right) ,
\eea
and stabilise itself by attracting neighbouring segments. We call such classical solutions the ``toroid states.''
Whenever $a$ increases and passes through the point
$\frac{2\pi n}{L}$ for $n\in\Z_+$, another toroid state
appears with the increased winding number $n$. Note that the radius of
the toroid state is given by $\frac{1}{a}$. 
When the chain of contour length $L$ winds $N(L)$ times we have the
$N(L)$ circles of each length $\frac{2\pi}{a}$ and the rest $\left(
L-\frac{2\pi}{a}N(L) \right)$. The second and third terms in the second line of
eq.(\ref{Hamilwithattraction}) result from the former and the latter
respectively. 
When $0<a\leq \frac{2\pi}{L}$, the chain cannot
wind like the toroid states. Both ends of the chain are not
connected to each other, thus can move freely as well as any other
parts of the chain fluctuate. 
As long as the total energy of the chain does not exceed the bending energy of $\frac{2\pi^2 l}{L}$ at {}$a=\frac{2\pi}{L}$, 
they can whip with zero winding number. 
We call such low-energy states the ``whip states.'' 
Although the definition includes fluctuations around the classical solutions, unless otherwise stated, 
we primarily 
refer to the classical solutions of such states, which 
are rather bowstrings than whips.

We now explore the exact energy levels of the whip and toroid states, and
discuss the phase transitions between these states. The dimensionless Hamiltonian of the second classical solution
(\ref{classical sol}) is a function of $l, L, W$ and $a$: 
\bea
&{\cal H}_{cl}&\!\!\!(a,l,L,W)
\nn
&\equiv&\!\!\!\!\!\frac{Ll}{2} a^2\!+\!\frac{\pi W }{a} N(L) \left(N(L)+1\right)\!-\!W L\cdot N(L).\label{Hamiltoniancl}
\eea
This matches with the first classical solution when $\frac{N(L)}{a}=0$ for $a=0$ is defined. 
Accordingly, 
the above expression is valid for all classical solutions. 

\begin{figure}[h]
\centering
\includegraphics[width=8.5cm]{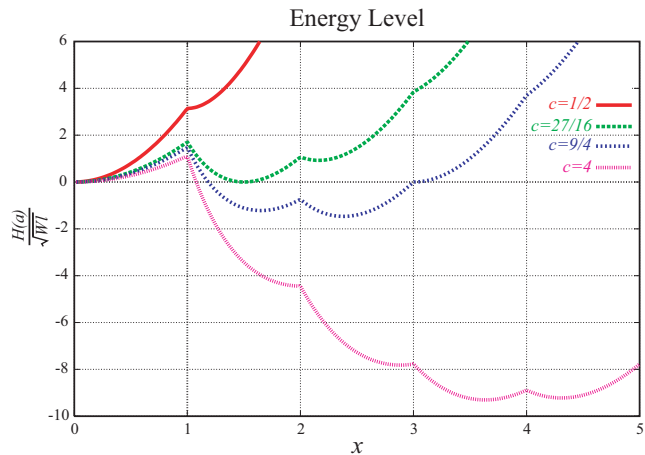} 
 \caption{The dependence of the energy ${\cal H}(a)$ on $x=a L/2\pi$ and $c$. ${\cal H}(a)$ is scaled by the factor of $\sqrt{Wl}$ for convenience. The definition of $c$ will appear shortly.}
      \label{fig:energy level}
\end{figure}

Consider first a case with $L$, $W$, and $l$ fixed.
By definition, 
${\cal H}(a) \equiv {\cal H}_{cl}(a,l,L,W)$ is continuous in the entire region of $a\geq 0$ and is a smooth function in each segment: $a\in \left[ \frac{2\pi N}{L}, \frac{2\pi(N+1)}{L} \right]$ for $N\in \Z_{\geq 0}$.
However, it is not
smooth at each joint of the segments: $\frac{aL}{2\pi} \in \Z_+$. We
plot, in Fig.\ref{fig:energy level}, the energy level as a function of
$a$ for different values of $c$, showing qualitative agreement with Conwell et al. \cite{CVH03}.
Given $N(L)=N$ is fixed, the Hamiltonian (\ref{Hamiltoniancl}) takes a
minimum at $a=a_c(N)\equiv\left( \frac{\pi W}{Ll}N(N+1)
\right)^{1/3}$. Accordingly, each segment falls into one of three cases: 
\begin{itemize}
\item[(i)] When $a_c(N) \leq \frac{2\pi N}{L}$, ${\cal H}(a)$ is a monotonic
function in the segment and takes its minimum at $a= \frac{2\pi N}{L}$. 
\item[(ii)] When $\frac{2\pi N}{L} <a_c(N)< \frac{2\pi (N+1)}{L}$, ${\cal H}(a)$ behaves quadratic in $a$ and takes its minimum at $a=a_c(N)$. 
\item[(iii)] When $\frac{2\pi (N+1)}{L}<a_c(N)$, ${\cal H}(a)$ is monotonic in the segment and takes its minimum at $a=\frac{2\pi (N+1)}{L}$.
\end{itemize}
The first and third cases are physically less relevant since they mean no (meta-) stable point in the segment. So, we focus on the second case.
Introducing a new parameter 
$c\equiv \left( \frac{L}{2\pi}\right)^{\!2}\! \frac{W}{2l}$,
the condition for $N$ or $a$ to be in the second case turns out to be 
\bea
N_L(c) < &N& < N_U(c)  \quad {\rm for~} c\geq 4,
\nn 
1\leq &N& < N_U(c)  \quad {\rm for~} 0\leq c < 4,
\eea
where $N = [aL/2\pi]$, 
$N_L(c) \equiv \frac{c}{2} \left( 1 - \frac{2}{c} +
\sqrt{1-\frac{4}{c}} \right)$, and $N_U(c) \equiv \frac{c}{2} \left( 1
+ \sqrt{1+\frac{4}{c}} \right)$. 

When $N_U(c) \leq 1$ (i.e. $c \leq \frac12$), 
the above second condition vanishes
and thus the whip states only survive at low energy. 
Therefore, at the critical value of $c=\frac12$, the whip phase to
whip-toroid co-existence phase transition may occur. On the other
hand, when $c > \frac12$, there always exists at least one (meta-)
stable state for some non-zero value of $a$ with positive winding
number $N(L)$. The number of minima are roughly given by the width of the region for $N$, i.e., $N_U(c) - N_L(c)$. 
For example, when $c \geq 4$, $N_U(c) - N_L(c)>3$, 
and therefore there are at least three minima with positive winding numbers greater than 1. 
When $0 < c < 4$, the condition of having three minima is $c>\frac94$. 
To summarise, when $c>\frac94$ there exist at least three minima with 
different winding numbers. 
This occurence of the multiple local minima is because the energy is given by the balance between the bending energy and the attractive potential energy. The former is a monotonically increasing function of $a$ while the latter is monotonically decreasing but not smooth function of $a$. 
This non-smoothness is a source of multiple stable states. 
Note that our precise analysis shows that the number of minima could
be reduced in some cases, 
for example, one can confirm it shortly by plotting the Hamiltonian (\ref{modH}) 
with some finite size effect on. 
The existence of multiple minima indicates the first order phase transitions between these stable states. 
In fact, 
when the energy of the $N=1$ stable toroid state becomes zero, 
the whip dominant to toroid dominant phase transition occurs. 
Such a value of $c$ is $27/16$. 
When $c>27/16$, the toroid states will dominate the action. 
Further discussions on the phase transitions can be seen by changing the value of $c$ \cite{IK06}.

One can plot the critical value of $c$ where the minimum of
the $N$-th segment emerges and vanishes. The lower bound of the $N$-th
segment is $c_L^{(N)} \!\!=\! \frac{N^2}{N+1}$, while the upper bound is
$c_U^{(N)} \!\!=\! \frac{(N+1)^2}{N}$. So, when $c_L^{(N)}\!\!< c < c_U^{(N)}$, 
the $N$-th segment has a minimal
and (meta-) stable point. For example, when $\frac12 < c < 4$, 
the first segment $a\in [\frac{2\pi}{L}, \frac{4\pi}{L}]$ (i.e. $N=1$) 
has a minimal point at $a=a_c(1)$. 

So far we have dealt with the classical solutions, which are derived
from the first derivative of the action. Thus, they may correspond
to the global/local minima of the action in the configuration space. However, 
the solutions are not necessarily stable 
unless we take into account the attraction, 
since the second derivative of the
action is zero or even negative in many cases. 
When we take the attraction into account, we observe the followings. 
When $c\geq 4$, the toroid states become stable under the quantum fluctuations,
since deviations from such toroid states will cost a large penalty in energy. 
When $c$ is small, 
a number of the whip states become equally or more probable to any toroid state.


For large $c$, the ground state --- the dominant toroid state of the winding number $N_c$ can be estimated by the inequality relation of $c$: $c_L^{(N)}< c <c_U^{(N)}$. It reads $N_c \simeq c$. 
Using this, we can estimate that the radius of our ideal toroid behaves $r_c = \frac{L}{2\pi N_c} = \frac{4\pi l}{W L}$.
Our ideal toroids have zero thickness, 
but the finite size effect of their cross sections can be approximated 
by the hexagonally arranged DNA chains with a van der Waals type interaction, i.e., with the effective nearest neighbour interactions. 
In the case of $N(L)\geq 4$,
it leads to the modified Hamiltonian:
\bea
{\cal H}(a)
 &=& \frac{Ll}{2}a^2 - \frac{2\pi W}{a} \VD(N(L))
\nn&&
- \frac{2\pi W}{a} \! \left(\frac{aL}{2\pi} -N(L) \right) Gap(N(L)), 
\label{modH}
\eea
where $\VD (N) \equiv 3 N -2 \sqrt{3} \sqrt{N-1/4}$ and $Gap(N)\equiv \VD (N+1)- \VD (N)$.
Following the same procedure for $c_L^{(N)}< c <c_U^{(N)}$, one can find 
$
 N_c \simeq \left( 2\sqrt{3} c \right)^{\frac{2}{5}} 
\label{N_c hex}
$
for large $c$. 
By $r_c\equiv \frac{L}{2\pi N_c}$, we now estimate the mean radius of the toroid (i.e. the average of inner and outer radii) in a physical
system in more detail. 
A coupling constant can be given by
$W\!=\!\frac{1}{l_m}\!\!\left(\frac{k\epsilon}{k_BT}\right)$ where $k$
is the number of the electric dipoles in a monomer segment, which create van der Waals interactions of the magnitude $\epsilon$. $l_m$ denotes the length of the monomer along the chain contour,
taken to be $l_m\simeq5\,bp=1.66\,nm$ in the end.
Substituting $N_c$ and the above, we obtain
$$
r_c \simeq 
 {\left(6\pi\right)}^{-\frac15}L^{\frac15}
  {\left(\frac{l}{W}\right)}^{\!\!\frac25}
 ={\left(6\pi\right)}^{-\frac15}L^{\frac15}
  {\left(l_ml\right)}^{\frac25}
  {\left(\frac{k\epsilon}{k_BT}\right)}^{\!\!-\frac25}.
$$
This $L^{\frac15}$ dependence agrees with the proposed exponent in the
asymptotic limit \cite{SIGPB03,SGM02,PW00,MKPW05}. 

We estimate the mean toroidal radius of T$4$ DNA in low ionic
conditions reported in ref.\cite{YYK98}. 
Using $L=57{\mu}m$, $l\,{\simeq}\,50\!\sim\!60\,nm$, 
and $l_m$, the mean radius
of the toroid is $r_c = 29.09 B^{-\frac25} \sim 31.29 B^{-\frac25}$ $[nm]$, 
where $B\equiv \frac{k\epsilon}{k_BT}$. 
This result is in good agreement with the experiment $r_c \simeq 28.5$ $nm$
for $B \,{\sim}\, 1.15$. 
The same argument for Sperm DNA packaged by protamines ($L=20.4{\mu}m$)
gives the analytic value $r_c = 23.69 B^{-\frac25} \sim 25.48 B^{-\frac25}$ $[nm]$, which is
comparable to an experimental result $r_c{\simeq}26.25$ $nm$. 
Note that the latter has a larger diameter of the segment and is expected to have the weaker interaction with smaller $B$.

The exponents $\nu$ predicted in the literature for
$r_c \sim L^{\nu}$ are $\nu=\frac15$ in most cases 
\cite{SIGPB03,SGM02,PW00,MKPW05}.
However, they are inconsistent with the experimentally well
known observation that the radius is independent of the chain length 
\cite{B91,B96,YYK98}. 
This might suggest that the real interaction is not van der
Waals like, or at least is not a single van der Waals type interaction.
It should be noted here that combinations of our ideal toroid and its finite size effect can give a range of $\nu = -1 \sim \frac15$ in some region.
Another important and interesting remark is that, in fact, when we apply Coulomb like interactions to our approximation,
we observe some 
asymptotic behaviour that the radius remains constant as $L$ changes. The precise analysis is to be studied and given in the near future.

\vspace*{10pt}
\noindent
{\bf Acknowledgments}

{Y.I. is grateful to K. Nagayama for his
discussions and encouragement. N.K. is grateful to N. Uchida and T. Kawakatsu
for stimulating discussions, and to T. A. Vilgis for his earlier
discussions of the field theory of globule-toroid transition which
led us to this direction. N.K. acknowledges the Deutsche Forschungsgemeinschaft for financial support.}

\end{document}